\documentclass[aps,prl,showpacs,twocolumn,amssymb]{revtex4}
\usepackage{graphicx}
\begin{document}

\draft

\title{Sliding charge-density-wave in two-dimensional rare-earth tellurides}

\author{A.~A.~Sinchenko},
\address{National research nuclear university (MEPhI), 115409 Moscow,
Russia}

\author{P.~Lejay}
\address{Institut NEEL, CNRS and Universit\'{e} Joseph Fourier, BP 166,
38042 Grenoble, France}

\author{P.~Monceau},
\address{Institut NEEL, CNRS and Universit\'{e} Joseph Fourier, BP 166,
38042 Grenoble, France}

\date{\today}

\begin{abstract}

Nonlinear transport properties are reported in the layered DyTe$_3$
compound at temperature below the charge-density-wave (CDW)
transition, $T_P=302$ K. Conductivity is increasing sharply above
the threshold electric field. Under application of a rf field
Shapiro steps are clearly observed. These features demonstrate for
the first time CDW sliding in two-dimensional compounds.

\end{abstract}

\maketitle

Interaction between pairs of quasiparticles often leads to
broken-symmetry ground states in solids. Typical examples are the
formation of Cooper pairs in superconductors, and charge (CDW) and
spin (SDW) density waves driven by electron-phonon and
electron-electron interactions respectively \cite{Gruner}. Density
wave formation is favoured by nesting of parallel Fermi surface (FS)
at $+k_F$ and $-k_F$. The first canonical example was found fifty
years ago when the antiferromagnetism in Cr was identified as an
incommensurate SDW transition \cite{Overhauser62} resulting from
nesting of electron-hole pockets sheets having similar shape
\cite{Lomer62} (for a review see also \cite{Fawcett88,Fawcett94}).

A CDW ground state is characterized by a concomitant spatial
modulation $\sim \cos(Qx+\varphi)$ of the electron density and a
periodic lattice distortion with the same $Q_{CDW}=2k_F$ wave vector
inducing opening of a gap in the electron spectrum. The coupled
electron-phonon which leads to the Kohn anomaly at high temperature
is split into two different modes below the Peierls transition: on
optical mode (called amplitudon) and on acoustic phase mode, or
phason \cite{Gruner}.

For ideal 1D conductors, the phase mode exists at zero frequency at
$Q_{CDW}$ realizing the Goldstone mode in systems with spontaneously
broken continuous symmetry. Then, a persistent current is provided
by the sliding motion of the CDW \cite{Frohlich54}. However various
mechanisms such as impurities, interchain interaction or
commensurability pin the phase of the CDW and introduce a gap in the
excitation of the phase mode which prevents the dc Fr\"{o}hlich
conductivity but leads to a large low frequency ac conductivity. The
pinning energy must be overcome to initiate the CDW sliding; that
can be achieved by the application of an electric field of a
sufficient strength \cite{Gruner}. In addition to the increase of
conductivity associated with collective CDW motion above a threshold
electric field, $E_T$, a periodic time dependent voltage is
generated as well as a broad band noise \cite{Gruner}. Sliding CDW
properties have been observed in inorganic NbSe$_3$, TaS$_3$,
K$_{0.3}$MoO$_3$, (TaSe$_4$)$_2$I as well as in organic (TTF-TCNQ),
(TMTSF)$_2$X, (Per)$_2$M(Mnt)$_2$ and (fluoranthene)$_2$X
one-dimensional compounds \cite{Monceau12}.

Attempts to detect sliding effect in 2D systems are, up to now,
unsuccessful. Thus, absence of nonlinearity in current-voltage
characteristics (IVc) was reported in 2H- TaSe$_2$ and 1T-TaSe$_2$
up to electric field 1 V/cm and 10 V/cm respectively
\cite{DiSalvo80}. These values are much higher that typical $E_T$ in
1D compounds. One possible reason might be a commensurability
pinning resulting from the triple-$\bf Q$ structure with three wave
vectors of equal amplitude, $120^\circ$ apart.

Sliding properties were also searched in other 2D structures. Thus,
2D quantum systems of electrons of extremely low disorder in
GaAs/AlGaAs heterejunctions were shown to exhibit conduction typical
of pinned CDWs \cite{Goldman90}. Microwave resonance in conductivity
were understood as the pinning mode of the electron or charge
ordered crystal \cite{Chen03,Sambandamurthy08}. For reentrant
integer quantum Hall states, a sharp threshold and a periodic
voltage associated with broad band noise were measured
\cite{Cooper03}. Although such effects are very similar to
characteristic features of CDW sliding, the very low frequency of
the ac voltage generated in the non-linear state disregards the
analogy with a simple CDW model.

It was recently claimed that the superstructure in the stripe phase
of a thin film of La$_{0.5}$Ca$_{0.5}$MnO$_3$ was a prototype CDW
with collective transport properties \cite{Cox08}. This statement
was based on observation of hysteresis in IVc and broad band noise.
However, no clear threshold field or ac voltage generation was
observed, that was explained by a large impurity density. Broad band
noise observation in manganites was also reported in
\cite{Wahl03,Barone09}. Nonlinear transport properties in manganites
were challenged in \cite{Fisher10}.

Recently a new class of layered compounds, namely rare-earth
tritellurides $R$Te$_3$ ($R$ =Y, La, Ce, Nd, Sm, Gd, Tb, Ho, Dy, Er,
Tm), has raised an intense research activity on CDW
\cite{DiMasi95,Brouet08,Ru08}. These systems exhibit an
incommensurate CDW through the whole $R$ series with a wave vector
${\bf Q}_{CDW1}=(0,0,\sim2/7 c^*)$ with a Peierls transition
temperature above 300 K for the light atoms (La, Ce, Nd). For the
heavier $R$ (Dy, Ho, Er, Tm) a second CDW occurs with the wave
vector ${\bf Q}_{CDW2}=(\sim2/7 a^*,0,0)$. The superlattice peaks
measured from X-ray diffraction are very sharp and indicate a long
range 3D CDW order \cite{Ru08}.

All $R$Te$_3$ compounds have the same orthorhombic structure
($Cmcm$) in the unmodulated state. The structure is formed of blocks
of $[R_2^{3+}$Te$_2^{2-}]^+$ layers sandwiched between the Te$^-$
layers and stacked together along the $b$-axis with weak van der
Waals gap ($\sim3.8${\AA}) between them. In the ($a,c$) plane Te--Te
distances are 3.1 {\AA} (to be compared to the covalent Te--Te bond
of 2.8 {\AA}). The charge transfer from $R$ ions to Te square yields
the $p$-band for Te square sheets to be partially filled. Band
structure calculated with simple tight-binding approximation reveals
that the electronic bands of $E_F$ derive from $p_x$ and $p_z$ in
plane Te orbitals, leading to a simple FS. Splitting of the bands
occurs due to bilayer Te sheets in the unit cell shifted one with
respect to the other by $c/2$. The hopping between orbitals along a
given direction ($x$ or $z$), $t_\parallel$, is much larger than the
hopping between same orbitals on neigbour rows, $t_\perp$. In
addition hopping between second-neigbour, $t^\prime$, mixes the
$p_x$ and $p_z$ bands. Typically it was estimated
$t_\parallel\simeq2.0$ eV, $t_\perp\simeq0.37$ eV and
$t^\prime\simeq0.16$ eV \cite{Yao06}. The finite value of $t_\perp$
introduce the warping of FS; $t_\perp$ is much larger than 300 K,
indicating that 1D approximations are not applicable to $R$Te$_3$
compounds.

Amplitude CDW excitations in $R$Te$_3$ were probed by Raman
scattering \cite{Lavagnini08,Lavagnini10} and femtosecond pump-probe
spectroscopy \cite{Yusupov10}. On the other hand, collective charge
phase excitations could not be observed in far-infrared measurements
\cite{Lavagnini09} due to screening by the residual metallic
component at the Fermi surface. But the phase collective mode is
accessible through nonlinear transport properties as we report
hereafter in DyTe$_3$. It is the first evidence of CDW sliding in a
quasi-2D systems.

DyTe$_3$ was chosen because in this compound the CDW appears just at
room temperature at $T_{CDW1}=302$ K and the lower CDW at
$T_{CDW2}=49$ K is the largest in the $R$Te$_3$ series. The
component of the wavevector ${\bf Q}_1$ along $c^*$ was found to be
0.2984 \cite{Malliakas08} (or 0.7061 according to whether the 2D
Brillouin zone for a single Te plane or the 3D Brillouin zone for
the unit cell is considered). This value is slightly far away from
2/7 or 5/7 (0.286 or 0.714) estimated from band filling resulting
from charge transfer between $R$ ions and Te sheets.

A slightly modified method described in \cite{Ru06} has been used
for the growth of single crystal of DyTe$_3$: single crystals were
grown by a self-flux technique under purified argon atmosphere in a
sealed quartz tube. High quality starting elements were used: 4N, Te
6N and the molar composition of the binary system was: (DyTe$_3$ +
90 Te). The mixture has been heated, in a same run, at 550$^\circ$C
and 850$^\circ$C respectively for 2 days before cooling down at with
a ramping of $2^\circ/$hour to 450$^\circ$C and then quenched in air
to room temperature. The millimeter scale plate-like single crystals
were mechanically extracted from the bulk. The $b$-axis has been
checked to be perpendicular to the large surface of crystals and its
value refined $b=25.420$ {\AA} for DyTe$_3$.

Thin single-crystal samples were prepared by micromechanical
exfoliation of relatively thick crystals glued on a sapphire
substrate. From homogeneous square single crystals with thickness
$0.4\div5.0$ $\mu$m, we cut bridges with width $50\div100$ $\mu$m.
Measurements of current-voltage characteristics (IVs) and their
derivatives have been performed with a conventional 4-probe
configuration. Contacts were prepared from In by cold soldering. The
distance between potential probes varied in the range $0.2\div0.5$
mm. For studying nonstationary effects a radiofrequency (rf) current
was superposed on the dc current using current contacts connected
with the generator via two capacitors.

\begin{figure}[t]
\includegraphics[width=8cm]{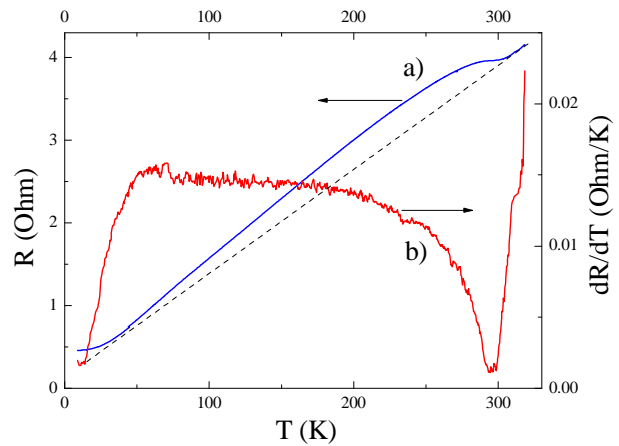}
\caption{\label{F1}(color online) Temperature dependence of
resistance (curve a) and its derivative (curve b) in the (a--c)
plane of a DyTe$_3$ single crystal. The dotted line is the linear
extrapolation of the resistance from above $T_{CDW1}$.}
\end{figure}

Fig.\ref{F1} shows the characteristic $R(T)$ (a) and $dR(T)/dT$ (b)
dependencies in the (a--c) plane for one of the DyTe$_3$ sample we
measured. The small increase in resistance below $T_{CDW1}=302$ K is
the signature of the Peierls transition in this compound. On the
$dR(T)/dT$ this effect is much more pronounced. Our data in
Fig.\ref{F1} are very similar to those in \cite{Ru08}. No visible
variation of the resistance was detectable for the CDW transition at
$T_{CDW2}$.

\begin{figure}[t]
\includegraphics[width=8cm]{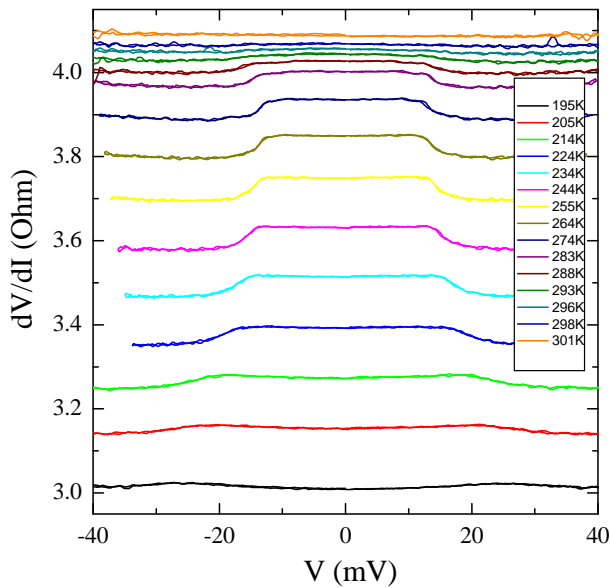}
\caption{\label{F2}(color online) Differential resistance
$R_d=dV/dI$ as a function of the applied voltage $V$ at different
temperatures varied from 195 K to 301 K for a DyTe$_3$ single
crystal with a thickness 0.5 $\mu$m.}
\end{figure}

In Fig.\ref{F2} we have drawn the differential IVc in the
temperature range $195<T<301$ K for a sample with a thickness of 0.5
$\mu$m. A pronounced non-linearity is observed below 300 K. Although
the amplitude of this non-linearity is small ($\Delta
R/R\sim10^{-2}$), the threshold behavior typical for the transition
to the sliding CDW state is clearly seen: the IVs are Ohmic for
voltage less than some threshold voltage $V_t$, and for voltages in
excess of this value the differential resistance, $R_d=dV/dI$,
decreases sharply. At temperatures $T>T_P=302$ K the IVs are Ohmic
without any non-linearities except very little Joule heating which
appears as a small parabolic increase of the differential
resistance. For thicker samples, the threshold behavior is smeared
by Joule heating.

\begin{figure}[t]
\includegraphics[width=8cm]{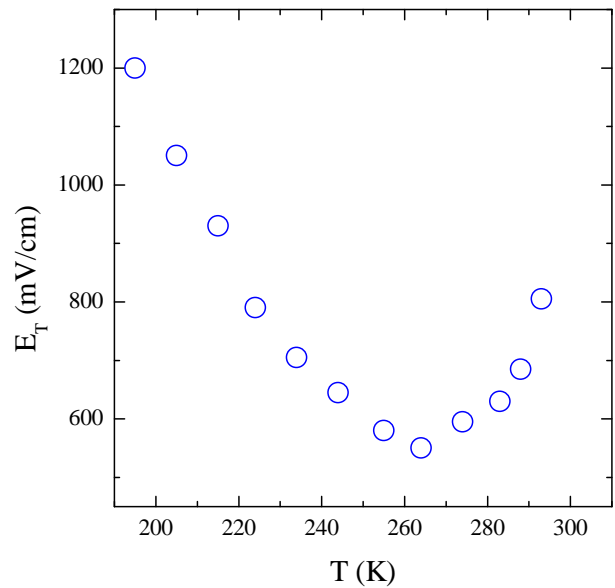}
\caption{\label{F3}(color online) Temperature dependence of the
threshold field, $E_T$, for the same sample as shown in Fig.2.}
\end{figure}

The very sharp onset of the observed non-linearity allows to
determine the temperature dependence of the threshold electric
field, $E_T$, which is shown in Fig.\ref{F3}. As can be seen, $E_T$
initially decreases in the temperature range $301\div265$ K and
increases when the temperature is decreased further. Note that such
a behavior is typical for $E_T$ in quasi-one dimensional systems
with a CDW \cite{Gruner}.

\begin{figure}[t]
\includegraphics[width=7.5cm]{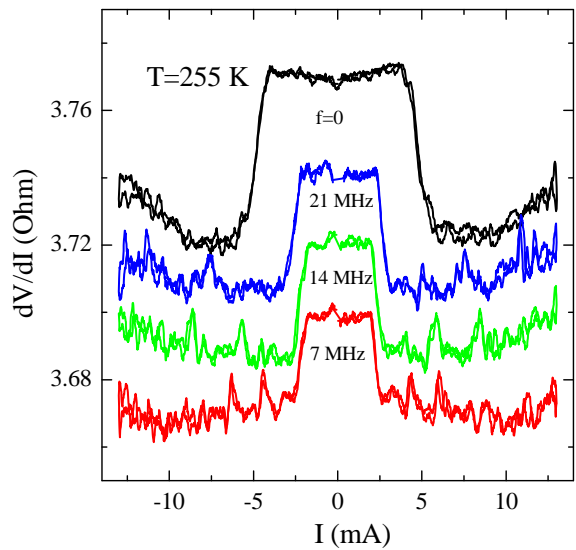}
\caption{\label{F4}(color online) $dV/dI(I)$ dependencies at $T=255$
K under application of a rf field with a given frequency 7, 14 and
21 MHz for the same sample as shown in Fig.2. For comparison, the
static (without rf field) differential curve is also shown. The
curves are shifted relatively to each other for clarity.}
\end{figure}

It is well known that the joint application of dc and rf driving
fields leads to appearance of harmonic and subharmonic Shapiro steps
in the dc IV characteristics of 1D CDWs \cite{Gruner}. In the
present work we have also observed Shapiro steps at such
experimental conditions. Fig.\ref{F4} shows $dV/dI(I)$ dependencies
at $T=255$ K under application of a rf field with a frequency of 7,
14 and 21 MHz with the same rf-power. For comparison, the static
(without rf field) differential IV measured at this temperature is
also shown. The curves are shifted relatively to each over for
clarity.  First of all, note that application of a rf electric field
leads to reduction of the threshold electric field $E_T$. At the
same time, Shapiro steps appear in the $dV/dI(I)$ characteristic as
a sharp maxima in the differential resistance. With increasing
frequency the distance between neighbouring maxima increases
proportionally to the frequency.

The observation of Shapiro steps is one of the characteristic
features of a CDW collective transport \cite{Gruner}. Thus, the
observed non-linearities of dc IVc together with the observation of
Shapiro steps evidently indicate that a contribution to the
electrical transport from the collective CDW motion takes place in
the 2D CDW DyTe$_3$ compound.

It is worth to note that the effect of CDW on the resistance in the
(a--c) plane is very weak (see Fig.1) and consists only a few \% of
increase below $T_{CDW1}$. In fact the larger effect is along the
$b$-axis perpendicularly to the Te sheets \cite{Ru08}. It is
commonly explained that the reduction of the Fermi surface (FS) by
opening of the CDW gap reduces the scattering of electrons and
therefore the conductivity. From angle-resolved photoemission
(ARPES) and optical studies it was shown that at low temperature
about 30\% -40\% of FS in ErTe$_3$ and HoTe$_3$ is affected by
formation of the CDW \cite{Pfuner10}. The small decrease of
conductivity below $T_{CDW1}$ may indicate that electrical transport
does not significantly involve bands on which the CDWs occur.

Although the temperature dependence of the gap amplitude was found
to follow a BCS type, the ratio $2\Delta/k_BT_{CDW}\simeq15\div17$
is much larger than the mean field 3.52 value. Such a large
magnitude was already measured in transition metal dichalcogenides,
and specifically in 2H-TaSe$_2$. That led McMillan to develop a
strong coupling model for systems with short coherence length where
phonon frequencies are modified over large parts of reciprocal space
\cite{McMillan77}. The strong electron coupling is also revealed by
the stretching of the lattice along the $c$-axis resulting from the
CDW formation \cite{Ru08}.

In conclusion we have shown that the phase of the 3D long range
order CDW in the layered DyTe$_3$ compound can easily slide above a
small threshold field. We demonstrate thus, for the first time, CDW
sliding in two-dimensional compounds.

\acknowledgements

The authors are thankful to J. Marcus for help in the sample
preparation. The work has been supported by Russian State Fund for
the Basic Research (No. 11-02-01379), and partially performed in the
frame of the CNRS-RAS Associated International Laboratory between
Institute Neel and IRE "Physical properties of coherent electronic
states in coherent matter".

\end{document}